\documentclass[prd,
superscriptaddress,showpacs,nofootinbib,preprintnumbers]{revtex4}

\usepackage{amsmath}
\usepackage{amsfonts}
\usepackage{graphicx}
\usepackage{dcolumn}
\usepackage{hyperref}

\newcommand{\be}{\begin{equation}}
\newcommand{\ee}{\end{equation}}
\newcommand{\bea}{\begin{eqnarray}}
\newcommand{\eea}{\end{eqnarray}}
\newcommand{\beaa}{\begin{eqnarray*}}
\newcommand{\eeaa}{\end{eqnarray*}}

\newcommand{\nn}{\nonumber \\}

\newcommand{\e}{{\rm e}}

\begin{document}

\tolerance=5000

\title{Multiple Lambda cosmology: dark fluid with time-dependent
equation of state as classical analog of cosmological landscape}
\author{Shin'ichi Nojiri}
\email{nojiri@phys.nagoya-u.ac.jp}
\affiliation{Department of Physics, Nagoya University, Nagoya 464-8602, Japan}
\author{Sergei D. Odintsov\footnote{also at Lab. Fundam. Study, Tomsk State
Pedagogical University, Tomsk}}
\email{odintsov@ieec.uab.es}
\affiliation{Instituci\`{o} Catalana de Recerca i Estudis Avan\c{c}ats (ICREA)
and Institut de Ciencies de l'Espai (IEEC-CSIC),
Campus UAB, Facultat de Ciencies, Torre C5-Par-2a pl, E-08193 Bellaterra
(Barcelona), Spain}

\begin{abstract}

We discuss FRW universe with time-dependent EoS dark fluid which leads to
multiple de Sitter space. This model (as well as its scalar-tensor
version)
may be considered as some classical analog of cosmological landscape.
The universe expansion history may look as transitions between different
deSitter eras which suggests the interesting solution for
cosmological constant problem. For specific time-dependent EoS dark fluid
the possibility of transitions between universe regions with positive and
negative cosmological constant is also established.
\end{abstract}


\maketitle

\noindent
1. In the study of celebrated dark energy problem (for a review, see
\cite{sami})
the ideal fluid with specific (sometimes, strange) equation of state
(EoS) remains to be
the simplest possibility for the description of the current cosmic
acceleration. Various examples of ideal fluid EoS may be considered for
this
purpose: constant EoS with negative pressure, imperfect
EoS \cite{salvatore}, general EoS \cite{general}, inhomogeneous EoS
\cite{inh} (with time-dependent bulk viscosity as particular
case \cite{brevik}) and so on (for extensive review, see \cite{sami}).
Trying to save General Relativity and to explain the cosmic acceleration
at the same time may lead to the conjecture about even more exotic dark
fluids. Moreover, General Relativity with ideal fluid of any type may be
rewritten in the equivalent form \cite{capoplb} as modified gravity
(for general review, see \cite{review}). Of course, the introduction
of ideal fluid with
complicated equation of state is somehow phenomenological approach
because no explanation of the origin of such dark fluid follows.

The interesting possibility to explain the dark fluid origin may be related
with string theory (for recent review of some attempts in this direction,
see\cite{irina}). The following sequence may be proposed: string/M-theory
is approximated by modified (super)gravity \cite{review} which is observed
finally as General
Relativity with exotic dark fluid. If such conjecture is (even partially)
true, it is expected that some string-related phenomena may be typical
in dark energy universe. One celebrated stringy effect possibly related
with early universe is string landscape (see, for instance, \cite{land})
which may lead to some observational consequences (see, for instance
, \cite{mar}). The cosmological landscape may be responsible for discrete
mass spectrum of scalar/spinor equations \cite{igor}.

In the present Letter we suggest the dark fluid with time-dependent
EoS which may be considered as simple classical analog of string landscape.
It leads to multiple de Sitter space at the same moment.
It may also describe transitions between de Sitter universes, suggesting
the way to solve the cosmological constant problem.
Unlike to proper string landscape the true ground state may be still
selected because the on-shell action is not the same for all these
de Sitter solutions. Some modification of dark fluid EoS suggests
the possible transition between regions with positive and negative
cosmological constant.

\

\noindent
2. Let us study the specific model of ideal fluid which leads to multiple
Lambda cosmology. The FRW equations are given by
\be
\label{k3}
\frac{3}{\kappa^2}H^2 = \rho\ ,\quad
 - \frac{2}{\kappa^2}\dot H= p + \rho\ .
\ee
Here $\rho$ is the energy density and $p$ is the pressure.
Instead of (\ref{k3}), one may include the cosmological constant from
gravity:
\be
\label{k3B}
\frac{3}{\kappa^2}H^2 = \rho+ \frac{\Lambda}{\kappa^2}\ ,\quad
 - \frac{2}{\kappa^2}\dot H= p + \rho\ .
\ee
We can, however, redefine $\rho$ and $p$ to absorb the contribution
 from the cosmological constant:
\be
\label{k3C}
\rho\to \rho - \frac{\Lambda}{\kappa^2}\ ,\quad
p\to p + \frac{\Lambda}{\kappa^2}\ .
\ee
By the redefinition (\ref{k3C}), we reobtain (\ref{k3}).
Hence, it is enough to consider only dark fluid in FRW equation.

If $\rho$ and $p$ are given in terms of the function $f(q)$ with a
parameter $q$ as (compare with similar Anzats \cite{grg})
\be
\label{kk1}
\rho = \frac{3}{\kappa^2}f(q)^2\ ,\quad
p=-\frac{3}{\kappa^2}f(q)^2 - \frac{2}{\kappa^2}f'(q)\ ,
\ee
 the following solution of Eq.(\ref{k3}) exists:
\be
\label{ML1}
H=f(t)\ .
\ee
Note that the origin of the time can be chosen arbitrary.
In (\ref{ML1}), $t=q$ but one may choose $t=q+t_0$ with an
arbitrary constant $t_0$.
This shows that besides the solution (\ref{ML1}),
$H=f(t-t_0)$ can be a solution.

If we delete $q$ in (\ref{kk1}), we have a general equation of state (EoS):
\be
\label{ML2}
F(\rho,p)=0\ .
\ee
In case that $f'(q)=0$ has a solution $q=q_0$, there is a solution
where $H$ is a constant:
\be
\label{ML3}
H=H_0\equiv f(q_0)\ ,
\ee
where $\rho=-p$, which corresponds to the effective cosmological constant.
Then if there is more than one solution satisfying $f'(q)=0$ as $q=q_n$,
$n=0,1,2, \cdots$,
the theory could have effectively several different cosmological
constants:
\be
\label{ML3b}
H=f(q_n)\ ,\quad \Lambda_n=3f(q_n)^2\ .
\ee
Note that the solutions (\ref{ML3b}) corresponding to
multiple cosmological constants may
exist at the same moment which shows some analogy with cosmological
landscape.
Let us assume there is a solution corresponding to $q_n$. By perturbation,
the solution may transit to
another solution, say $q_{n+1}$. Hence, the transition period is
proportional to $T_{n,n+1}=q_{n+q} - q_n$.

As an example, we consider the ideal fluid with the following
time-dependent EoS
\be
\label{ML4}
f(q)=f_0 - f_1 \e^{-\lambda q}\left(\frac{\cos(\omega q + \alpha)}{\sqrt{\lambda^2 +\omega^2}}
 - \frac{1}{\lambda}\right)\ .
\ee
Here $f_0$ and $f_1$ are constants and $f_0>f_1>0$ and $\alpha$ ($0<\alpha<\pi$, $\alpha\neq \pi/2$)
is defined by
\be
\label{ML5}
\tan\alpha=\frac{\omega}{\lambda}\ ,
\ee
that is
\be
\label{ML6}
\cos(\omega q + \alpha)=\frac{\lambda \cos \omega q
 - \omega \sin \omega q}{\sqrt{\lambda^2 +\omega^2}}\ .
\ee
Then one finds
\be
\label{ML6b}
f'(q)= f_1\e^{-\lambda q}\left(\cos\omega q - 1\right)\ .
\ee
Since $f'(q)=0$ when $q=\frac{2\pi n}{\omega}$ with an integer $n$,
we have effectively multiple
cosmological constant:
\be
\label{ML7}
\Lambda_n = 3\left\{f_0 - f_1 \e^{-2\pi n \lambda/\omega }
\left(\frac{\cos(\alpha)}{\sqrt{\lambda^2 +\omega^2}}
 - \frac{1}{\lambda}\right)\right\}^2
= 3\left\{f_0 + \frac{f_1 \e^{-2\pi n \lambda/\omega }\omega^2}{\lambda\left(\lambda^2
+ \omega^2\right)}\right\}^2\ .
\ee
Eq.(\ref{ML1}) shows also that there is a time dependent solution:
\be
\label{ML8}
H(t)=f(t)=f_0 - f_1 \e^{-\lambda t}\left(\frac{\cos(\omega t + \alpha)}{\sqrt{\lambda^2 +\omega^2}}
 - \frac{1}{\lambda}\right)\ .
\ee
When $t=2\pi n/\omega$, it follows $\rho=-p$ corresponding to the
cosmological constants in (\ref{ML7}).
Note that the solutions with cosmological constant (\ref{ML7}) may exist
at
the same moment. Such the universe could be considered as kind of
multiverse \cite{pedro} as it could be that some its regions may be
inaccessible like in the model \cite{brett}.
If we have a solution corresponding to $q_n$, by perturbation, the solution could transit to
another solution correponding to $q_{n+1}$. The time for transition could be given by $T=q_{n+q} - q_n=2\pi/\omega$.
Then the solution (\ref{ML8}) describes the transitions
 between different $\Lambda$ cosmologies (for related proposal of cascading universe, see \cite{scott}).
Since $\Lambda_n>\Lambda_{n+1}$ in (\ref{ML7}), the transition could occur
 from larger one to smaller one.
In the limit of $t\to \infty$, $H(t)$ goes to $f_0$, which corresponds to
\be
\label{ML8b}
\Lambda_\infty \equiv \lim_{n\to +\infty} \Lambda_n = 3f_0^2\ .
\ee
If $t\to \infty$ corresponds to the present time, we find $f_0\sim 10^{-33}$ eV.

As a second slightly different example, one may consider the following
ideal fluid:
\be
\label{MLa1}
f(q)=g_0 \e^{ - g_1 \left( q - \frac{1}{\omega} \sin \omega q \right)}\ ,
\ee
which gives
\be
\label{MLa2}
f'(q)= - g_0 g_1 \left(1 - \cos \omega q\right)
\e^{ - g_1 \left( 1 - \frac{1}{\omega} \sin \omega q \right)}\ .
\ee
In (\ref{MLa1}), it is assumed $g_0$, $g_1$, and $\omega$ are constants.
Therefore, $f'(q)=0$ when $q=\frac{2\pi n}{\omega}$ with an integer $n$,
again. The effective multiple cosmological constant occurs:
\be
\label{MLa3}
\Lambda_n = 3g_0^2 \e^{ - \frac{4\pi n g_1}{\omega}}\ ,
\ee
Eq.(\ref{MLa1}) shows that there exists also a time dependent solution:
\be
\label{MLa4}
H(t)= f(t)=g_0 \e^{ - g_1 \left( t - \frac{1}{\omega} \sin \omega t \right)}\ ,
\ee
Again $t=2\pi n/\omega$ corresponds to the cosmological constants in (\ref{MLa1}) and therefore
the time-dependent solution could describe the transition between the cosmological constants, from
larger one to smaller one. In the limit of $t\to +\infty$ or $n\to +\infty$, the effective cosmological
constant vanishes: $\lim_{n\to +\infty} \Lambda_n = 0$, which is different from
the example in (\ref{ML4}) as found in (\ref{ML8b}).

It may be shown that the EoS (\ref{ML2}) or (\ref{kk1}) can be
realized in the scalar-tensor theory,
whose action is given by
\bea
\label{k1}
&& S = \int d^4 x \sqrt{-g}\left\{\frac{1}{2\kappa^2}R
 - \frac{1}{2}\omega(\phi)\partial_\mu \phi \partial^\mu \phi - V(\phi)\right\}\ , \nn
&& \omega(\phi)=- \frac{2}{\kappa^2}f'(\phi)\ ,\quad
V(\phi)=\frac{1}{\kappa^2}\left(3f(\phi)^2 + f'(\phi)\right)\ .
\eea
For the example (\ref{ML4}), the corresponding scalar-tensor theory looks
like
\bea
\label{ML9}
\omega(\phi) &=& - \frac{2f_1}{\kappa^2} \e^{-\lambda \phi}\left(\cos\omega \phi - 1\right)\ ,\nn
V(\phi) &=& \frac{1}{\kappa^2} \left\{
3 \left(f_0 - f_1 \e^{-\lambda \phi}\left(\frac{\cos(\omega \phi + \alpha)}{\sqrt{\lambda^2 +\omega^2}}
 - \frac{1}{\lambda}\right)\right)^2
+ f_1 \e^{-\lambda \phi}\left(\cos\omega \phi - 1\right)\right\}\ ,
\eea
and for the example (\ref{MLa1}), it follows
\bea
\label{ML10}
\omega(\phi) &=& \frac{2g_0 g_1}{\kappa^2} \left(1 - \cos \omega \phi\right)
\e^{ - g_1 \left( 1 - \frac{1}{\omega} \sin \omega \phi \right)}\ , \nn
V(\phi) &=& \frac{1}{\kappa^2} \left\{
3 g_0^2 \e^{ - 2 g_1 \left( \phi - \frac{1}{\omega} \sin \omega \phi \right)}
 - g_0 g_1 \left(1 - \cos \omega q\right)
\e^{ - g_1 \left( 1 - \frac{1}{\omega} \sin \omega q \right)}\right\}\ .
\eea
The appearence of exponential potentials may indicate to some connection
with string theory. For other examples of
scalar-tensor theory with de Sitter solutions, see \cite{faraoni}.

Note that the solution (\ref{ML1}) is always also a solution of the
corresponding scalar-tensor
theory (\ref{k1}) but the solution (\ref{ML3}) is not always a solution of the scalar-tensor theory.
This is because the fact that energy conservation law $\dot\rho + 3H(\rho
+ p)=0$ is the first order differential
equation while the scalar field equation  is the
second order differential equation:
\be
\label{ML11}
0=\omega(\phi)\ddot \phi + \frac{1}{2}\omega'(\phi){\dot\phi}^2 + 3H\omega(\phi)\dot\phi + V'(\phi)\ .
\ee
Hence, even if the conservation law is satisfied, the field equation
(\ref{ML11}) is not always satisfied.
The solution (\ref{ML3}) requires that $\phi=q_0$ with $f'(q_0)=0$ and therefore $\phi$ is a constant.
Then in order that the solution (\ref{ML3}) is also a solution of the scalar-tensor theory, we should
require $V'(\phi)=0$ at $\phi=q_0$. Since
\be
\label{ML12}
V'(\phi)=\frac{1}{\kappa^2}\left\{6f(\phi)f'(\phi) + f''(\phi)\right\}\ ,
\ee
if
\be
\label{ML13}
f''(q_0)=0\ ,
\ee
$\phi=q_0$ is also a solution of the scalar-tensor theory.

For the model (\ref{ML4}) or (\ref{ML9}), we find
\be
\label{ML14}
f''(\phi)=f_1\e^{-\lambda \phi}\left\{ - \lambda \left(\cos \omega \phi - 1\right) - \omega\sin \omega\phi \right\}\ ,
\ee
which vanishes when $\phi=\frac{2\pi n}{\omega}$ with an integer $n$.
 Since $f'(\phi)$ also vanishes there,
$\phi=\frac{2\pi n}{\omega}$ is a solution of the scalar-tensor theory.
Thus, the scalar-tensor theory has multiple cosmological constant solution
\be
\label{ML15}
\frac{\Lambda_n}{\kappa^2} = V\left(2\pi n/\omega\right) = \frac{3}{\kappa^2}f\left(2\pi n/\omega\right)^2\ ,
\ee
which corresponds to (\ref{ML3b}).
The first FRW equation (\ref{k3}) shows that
\be
\label{ML15b}
R=12H^2=4\kappa^2\rho=4\kappa^2 V\left(2\pi n/\omega\right) = 4\Lambda_n\ ,
\ee
and therefore the on-shell Lagrangian density ${\cal L}$ (\ref{k1}) is
given by
\be
\label{ML15c}
{\cal L}=\frac{R}{2\kappa^2} - V\left(2\pi n/\omega\right) = \frac{\Lambda_n}{\kappa^2}
= \frac{3}{\kappa^2}\left\{f_0 + \frac{f_1 \e^{-2\pi n \lambda/\omega }\omega^2}{\lambda\left(\lambda^2
+ \omega^2\right)}\right\}^2\ .
\ee
Hence, the energy of the corresponding cosmological constant solution
changes with $n$, which is different from string landscape where millions
of de Sitter vacua may have the same energy.

For the model (\ref{MLa1}) or (\ref{ML10}), we find
\be
\label{ML16}
f''(\phi)= - g_0 g_1 \left\{ \omega \sin\omega \phi - g_1 \left( 1 - \cos \omega \phi \right)^2\right\}
\e^{ - g_1 \left( 1 - \frac{1}{\omega} \sin \omega q \right)}\ .
\ee
Since both of $f'(\phi)$ and $f''(\phi)$ vanish when $\phi=\frac{2\pi n}{\omega}$ with an integer $n$,
$\phi=\frac{2\pi n}{\omega}$ is also a solution of
the corresponding scalar-tensor theory. Then the
scalar-tensor theory has multiple cosmological constant solution
(\ref{MLa3}).
In this example, the corresponding on-shell Lagrangian density is given
by
\be
\label{ML17}
{\cal L} = \frac{3g_0^2}{\kappa^2} \e^{ - \frac{4\pi n g_1}{\omega}}\ .
\ee

\

\noindent
3. Generalizing above EoS, we consider the case that $f(q)$ is given by two
functions $g$ and $h$ in
the following form:
\be
\label{ML18}
f(q)=g\left(h(q) - \sin h(q)\right)\ .
\ee
Since
\bea
\label{ML19}
f'(q)&=&g'\left(h(q) - \sin h(q)\right)\left(1 - \cos h(q)\right)h'(q)\ ,\nn
f''(q)&=& g''\left(h(q) - \sin h(q)\right)\left(1 - \cos h(q)\right)^2\left(h'(q)\right)^2
+ g'\left(h(q) - \sin h(q)\right)\left(1 - \cos h(q)\right)h''(q) \nn
&& + g'\left(h(q) - \sin h(q)\right) \sin h(q) \left(h'(q)\right)^2\ ,
\eea
one finds $f'(q)=f''(q)=0$ when $h(q)=2\pi n$ ($n$ : integer) if $g'$,
$g''$, $h'$, and $h''$ are not singular.
Hence, EoS (\ref{kk1}) and the corresponding scalar-tensor theory
(\ref{k1}) contains the multiple de Sitter solution which
corresponds to the cosmological constant (\ref{MLa3})
when $h(q)=2\pi n$ ($n$ : integer):
\be
\label{ML19b}
\Lambda_n=3 \left(g\left(2\pi n \right)\right)^2\ .
\ee
Moreover, such ideal fluid as well as corresponding scalar-tensor theory
also leads to the time-dependent solution, whose Hubble rate
is given by
\be
\label{ML20}
H(t)=g\left(h(t) - \sin h(t)\right)\ .
\ee
Such (oscillating) universe passes through the de Sitter eras specified
by different
cosmological constants.

For instance, with the choice
\be
\label{ML21}
h(q)=2\pi N \tanh \omega q\ ,
\ee
we have $2N-1$ classical deSitter vacua, where $f'(q)=f''(q)=0$ or
$h(q)=2\pi n$ with integers $n$'s ($-N<n<N$),
between $-\infty<q<\infty$ and the limits $q\to\pm\infty$ also correspond
to de Sitter vacua.

The effective EoS parameter $w_{\rm eff}$ is defined by
\be
\label{ML22}
w_{\rm eff}=-1 - \frac{2\dot H}{3H^2}\ .
\ee
When $w_{\rm eff}<-1/3$, the universe is accelerating but when $w_{\rm
eff}>-1/3$, the universe is decelerating.
As an example of (\ref{ML18}), one can consider
\be
\label{ML23}
f(q)=\frac{H_0}{H_1 + h(q) - \sin h(q)}\ ,
\ee
where $h(q)$ is chosen as (\ref{ML21}). Assume $H_1>2\pi N$ then
$f(q)$ is not singular everywhere.
One can now investigate the solution (\ref{ML20}). Let define
$q_1$ by
\be
\label{ML24}
-2\pi\left(N-1\right) = h(q_1)=2\pi N \tanh \omega q_1\ .
\ee
Since $\dot H=0$ when $t=-\infty$ or $t=q_1$, $w_{\rm eff}=-1$ when $t=-\infty$ or $t=q_1$, that is,
the universe is tentatively de Sitter space and therefore, the universe is
accelerating.
Hence, there should be an extremum of $w_{\rm eff}$ when $-\infty < t <
q_1$.
 For the example (\ref{ML23}), one gets
\be
\label{ML25}
\dot w_{\rm eff}= - \frac{2}{3 H_0}\left\{ \left( 1 - \cos h(t) \right) h''(t) + \sin h(t)
\left( h'(t) \right)^2 \right\}\ .
\ee
At the extremum, where $\dot w_{\rm eff}=0$, it follows
\be
\label{ML26}
1 - \cos h(t) = - \frac{h'(t)}{h''(t)}\sin h(t)\ .
\ee
Let denote $t$ satisfying the condition (\ref{ML26}) by $t_e$,
($-\infty<t_e<t_1$). Note that $t_e$ does not depend on $H_0$.
When $t=t_0$, it follows
\be
\label{ML27}
w_{\rm eff}=-1 + \frac{2\left(1 - \cos h(t_e) \right) h'(t_e)}{H_0}
= - 1 + \frac{4\pi N \omega \left(1 - \cos h(t_e) \right) }{H_0 \cosh^2 \omega t_e}\ .
\ee
In (\ref{ML27}), the second term is positive if $N\omega/H_0>0$.
Since $t_e$ does not depend on $H_0$, we may choose $H_0$ to be small
enough. Then $w_{\rm eff}$ can be
larger than $-1/3$, that is, the universe can be decelerating.
Since the universe is accelerating when $t=-\infty$ or $t=t_1$, the
primordial universe, where $t\to -\infty$,
is accelerating but when $t\to t_e$, the universe decelerates, that is,
there occurs the transition
from the accelerating universe to the decelerating universe. After that,
when $t\to t_1$, the universe turns to
acceleration again, that is, there occurs the transition from deceleration
to acceleration.
After $t$ goes through $t_1$, the universe passes through several de
Sitter vacua and in the limit of $t\to + \infty$,
the universe tends to the final de Sitter space where
$H\to H_{\rm final} = H_0/\left(H_1 + 2\pi N\right)$. $H(t)$
is monotonically decreasing function of $t$, the value of
$H=H_{\rm final}$ is minimum.
The first period of the deceleration between $-\infty<t<t_1$ could be the
order of $1/\omega$. Then in order to obtain the
realistic model, we should choose $\omega$ to be the order of the age
of the universe,
that is, ten billion years or
$\omega\sim 10^{-33}$\,eV. When $t=t_1$, one has $H(t_1)=H_0/\left(H_1 -
2\pi\left(N-1\right)\right)$.
Since $H_1>2\pi N$, we find $H_1 - 2\pi\left(N-1\right) \sim {\cal O}(1)$
 and the magnitude of $H(t_1)$ is the order
of $H_0$ and therefore $H_0\sim 10^{-33}$\,eV, which corresponds
to the observed value of the Hubble rate in the present
universe. Since $H(-\infty)=H_0/\left(H_1 - 2\pi N\right)$,
if the Hubble rate in the inflationary era is large enough,
it follows $H_1 \sim 2\pi N$.
Hence, the ideal fluid with time-dependent EoS of special form motivated
by string landscape considerations may suggest the solution of the
cosmological constant problem. Of course, similar scenario may be extended
for scalar-tensor theory with corresponding potentials where also the
unification of cosmological epochs may be achieved (for one related
example, see \cite{plbno}).

\

\noindent
4. In the previous section, dark fluid describing the transitions
between de Sitter eras is discussed. In this section, we consider if
the transition between de Sitter and Anti-de Sitter space is possible in
the
similar framework. When cosmological constant is negative and there is no
matter,
there is no time evolution of the universe. One may include the matter
with constant EoS  parameter $w$, whose energy and the pressure are given by
\be
\label{ML28}
\rho_m=\rho_0 a^{-3(1+w)}\ ,\quad p_m=w\rho_0 a^{-3(1+w)}\ .
\ee
Then instead of (\ref{k3}), we find
\be
\label{ML29}
\frac{3}{\kappa^2}H^2 = \rho + \rho_m\ ,\quad
 - \frac{2}{\kappa^2}\dot H= p + \rho + p_m + \rho_m \ .
\ee
Hence, if $\rho$ and $p$ are given in terms of the function $g(q)$ with a
parameter $q$ as
\be
\label{ML30}
\rho = \frac{3}{\kappa^2}g'(q)^2 - \rho_0 a_0^{-3(1+w)}\e^{-3(1+w)g(q)} \ ,\quad
p=-\frac{3}{\kappa^2}f(q)^2 - \frac{2}{\kappa^2}f'(q) - w \rho_0 a_0^{-3(1+w)}\e^{-3(1+w)g(q)}\ ,
\ee
the following solution of (\ref{ML29}) exists:
\be
\label{ML31}
H=g'(t)\quad \left(a(t)=a_0\e^{g(t)}\right)\ .
\ee
Note that the correponding scalar theory is given by,
 instead of (\ref{ML9}),
\bea
\label{ML32}
\omega(\phi) &=& - \frac{2f_1}{\kappa^2} \e^{-\lambda \phi}\left(\cos\omega \phi - 1\right)
 - \frac{w+1}{2} g_0\e^{-3(1+w)g(\phi)} \ ,\nn
V(\phi) &=& \frac{1}{\kappa^2} \left\{
3 \left(f_0 - f_1 \e^{-\lambda \phi}\left(\frac{\cos(\omega \phi + \alpha)}{\sqrt{\lambda^2 +\omega^2}}
 - \frac{1}{\lambda}\right)\right)^2 + f_1 \e^{-\lambda \phi}\left(\cos\omega \phi - 1\right)\right\} \nn
&& + \frac{w-1}{2} g_0\e^{-3(1+w)g(\phi)}\ .
\eea
Here
\be
\label{ML33}
a_0=\left(\frac{\rho_0}{g_0}\right)^{\frac{1}{3(1+w)}}\ .
\ee

We now consider the following model:
\be
\label{ML34}
g(q)=\alpha(q) + \frac{2}{3(1+w)}\ln \sin \Omega q\ .
\ee
Here $\Omega$ is a constant and $\alpha(q)$ is a proper function
which satisfies the  conditions
\be
\label{ML35}
\alpha(0)=\alpha(\pi/\Omega)=\alpha'(0)=\alpha'(\pi/\Omega)=0\ .
\ee
Assume, when $q\sim 0$,
\be
\label{ML36}
\alpha(q)\sim \alpha_0 \Omega^2 q^2 \ ,
\ee
and when $q\sim \pi/\Omega$,
\be
\label{ML37}
\alpha(q)\sim \tilde\alpha_0 \left(\pi - \Omega q\right)^2\ .
\ee
One gets
\be
\label{ML38}
\rho=\frac{3}{\kappa^2}H^2 - \rho_m
= \frac{3}{\kappa^2}\left(\alpha'(t) + \frac{2\Omega}{3(1+w)}\cot \Omega t\right)^2
 - \rho_0 a_0^{-3(1+w)}\e^{-3(1+w)\alpha(t)}\sin^{-2}\Omega t\ .
\ee
We now choose the parameters to satisfy the following condition:
\be
\label{ML39}
\rho_0 a_0^{-3(1+w)} = \frac{4\Omega^2}{3(1+w)^2\kappa^2}\ .
\ee
Then in the limit $t\to 0$,  $\rho$ goes to some constant:
\be
\label{ML40}
\rho\to \frac{12\Omega^2}{\kappa^2(1+w)}\left(\alpha_0 - \frac{1}{9(1+w)}\right)\ ,
\ee
and also in the limit $t\to \pi/\Omega$, $\rho$ goes to the constant:
\be
\label{ML41}
\rho\to -\frac{4\Omega^2}{\kappa^2(1+w)}\left(\tilde \alpha_0 + \frac{1}{3(1+w)}\right)\ .
\ee
Hence, in the limits of $t\to 0,\ \pi/\Omega$, $\rho$ behaves as a
cosmological constant.
For example, if it is chosen
\be
\label{ML42}
\alpha_0 > \frac{1}{9(1+w)}\ ,\quad
\tilde \alpha_0 > - \frac{1}{3(1+w)}\ ,
\ee
$\rho$ behaves as a positive cosmological constant when $t\sim 0$ and
as a negative cosmological constant when $t\sim \pi/\Omega$.
Therefore, the dark fluid (\ref{ML34}) could  describe the
transition between universe regions with positive and negative
cosmological constants.

\

\noindent
5. In summary, we discussed FRW universe with
time-dependent EoS dark fluid as a toy model
which leads to multiple de Sitter epochs. This suggests some
classical analog of cosmological landscape. It remains to show
which stringy
compactification may be described as an effective ideal fluid of such
(or similar) form.

The possibility of transition between different de Sitter eras is
demonstrated in the universe expansion history. That may suggest the
reasonable approach to the solution of cosmological constant problem,
 explaining its effective decrease. In addition, the possibility of
transition between regions with positive and negative cosmological
constant is demonstrated. It is not difficult to generalize the above
study for the presence of standard CDM matter. In such a way,
the standard $\Lambda$CDM cosmology which is known to be consistent with
observational data (for recent review, see \cite{leandros}) easily
emerges with reasonable explanation of the smallness of current
cosmological constant.

\

\noindent
{\bf
Acknowledgements.}
The work by S.N. was supported in part by Monbusho grant  no.18549001
(Japan) and 21st Century COE program of Nagoya Univ. provided by
JSPS (15COEEG01)
and that by S.D.O. by the project FIS2005-01181 (MEC, Spain), by AGAUR
(Generalitat de Catalunya), contract 2005SGR-00790, by LRSS project
N4489.2006.02 and by RFBR grant 06-01-00609 (Russia).

\end{document}